\documentclass[aps,prb,amsmath,amssymb,reprint]{revtex4-1}
\usepackage{graphicx}
\usepackage{bm}
\usepackage{fancyref}
\usepackage{placeins} 
\usepackage{flafter}  
\usepackage{float}
\usepackage{soul}
\usepackage{color}
\makeatletter
\let\Hy@backout\@gobble
\makeatother
\usepackage[version=3]{mhchem}
\usepackage{xcolor}
\usepackage{titlesec}
\titlespacing*{\section}{0pt}{1.1\baselineskip}{\baselineskip}

\begin{document}

\author{D. Davidovikj$^1$}
\email{d.davidovikj@tudelft.nl}
\author{P. H. Scheepers$^1$}
\author{H. S. J. van der Zant$^1$}
\author{P. G. Steeneken$^{1,2}$}

\affiliation{$^1$Kavli Institute of Nanoscience, Delft University of Technology, Lorentzweg 1, 2628 CJ Delft, The Netherlands \\
$^2$Department of Precision and Microsystems Engineering, Delft University of Technology, Mekelweg 2, 2628 CD, Delft, The Netherlands} 
\title{Static capacitive pressure sensing using a single graphene drum}

\begin{abstract}
 
To realize nanomechanical graphene-based pressure and gas sensors, it is beneficial to have a method to electrically readout the static displacement of a suspended graphene membrane. Capacitive readout, typical in micro-electro-mechanical systems (MEMS), gets increasingly challenging as one starts shrinking the dimensions of these devices, since the expected responsivity of such devices is below 0.1 aF/Pa. To overcome the challenges of detecting small capacitance changes, we design an electrical readout device fabricated on top of an insulating quartz substrate, maximizing the contribution of the suspended membrane to the total capacitance of the device. The capacitance of the drum is further increased by reducing the gap size to 110 nm. Using external pressure load, we demonstrate successful detection of capacitance changes of a single graphene drum down to 50 aF, and pressure differences down to 25 mbar.

\end{abstract}
\maketitle

Nanomechanical devices from suspended graphene and other two-dimensional materials have been receiving growing interest in the past few years for their potential as sensitive pressure~\cite{smith13pressure,dolleman15,smith16,chen16,patel16} and gas~\cite{bunch12,wang15,dolleman16} sensors. To realize integrated, small and low-power devices, it is necessary to have all-electrical on-chip transduction schemes, in contrast to the currently often employed laser interferometry techniques for the readout of their dynamic motion and static deflection. 

Reports on electrical readout of graphene membrane nanomechanical devices have employed readout schemes based on electrical transconductance~\cite{chen09,patel16} and piezoresistivity~\cite{smith13pressure,smith16}. Both of these rely on the change in the conductance of the membrane as a function of deflection, which is then used to sense the motion of the membrane. Although these methods can be very sensitive, the graphene conductance can also be affected by variations in gas composition, humidity, light intensity and temperature. Moreover, the conductance is not only related to the deflection of the graphene membrane, but depends also on material parameters like the electron mobility and piezoresistive coefficients. These approaches therefore require calibration and a high degree of stability of the graphene and insensitivity to variations in its surroundings.\\
\begin{figure}[h]
	\includegraphics[width=0.5\textwidth]{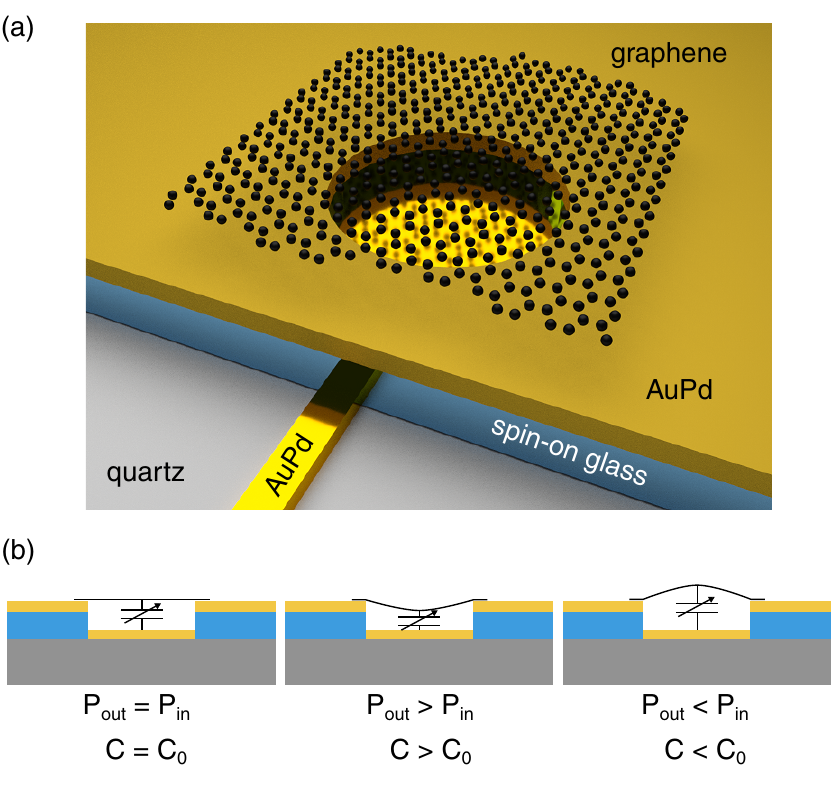}
	\caption{\label{fig:Fig1} (a) A 3D schematic of the device: a capacitor is formed between a graphene drum suspended over a metallic cavity and a bottom metallic electrode that runs underneath an insulating (spin-on glass) oxide layer. The entire device is fabricated on top of an insulating quartz wafer. (b) Actuation principle: external pressure load is applied. Depending on the pressure difference between the cavity and the outside environment, the nanodrum will bulge upwards or downwards resulting in a decrease or increase of the measured capacitance.}
\end{figure}
In contrast, the capacitance between a graphene membrane and a bottom electrode is, to first order, a function only of the geometry of the system and, therefore, the deflection of the membrane. A measurement of the capacitance of the membrane can therefore be used to calculate its deflection, which makes capacitance detection an interesting alternative method for electrical readout of nanomechanical graphene sensors. Dynamic (on-resonance) capacitive readout has been demonstrated on suspended graphene bridges~\cite{xu10,chen13thesis}. Measurements using static capacitive readout of the deflection of graphene membranes have been conducted on a voltage tunable capacitor array comprised of thousands of graphene membranes in parallel~\cite{abdelghany16}. 

Here we extend on this work by capacitive detection of the deflection of a \textit{single} graphene drum and demonstrating its performance as a pressure sensor.\\
\begin{figure*}[ht!]
	\includegraphics[width=\textwidth]{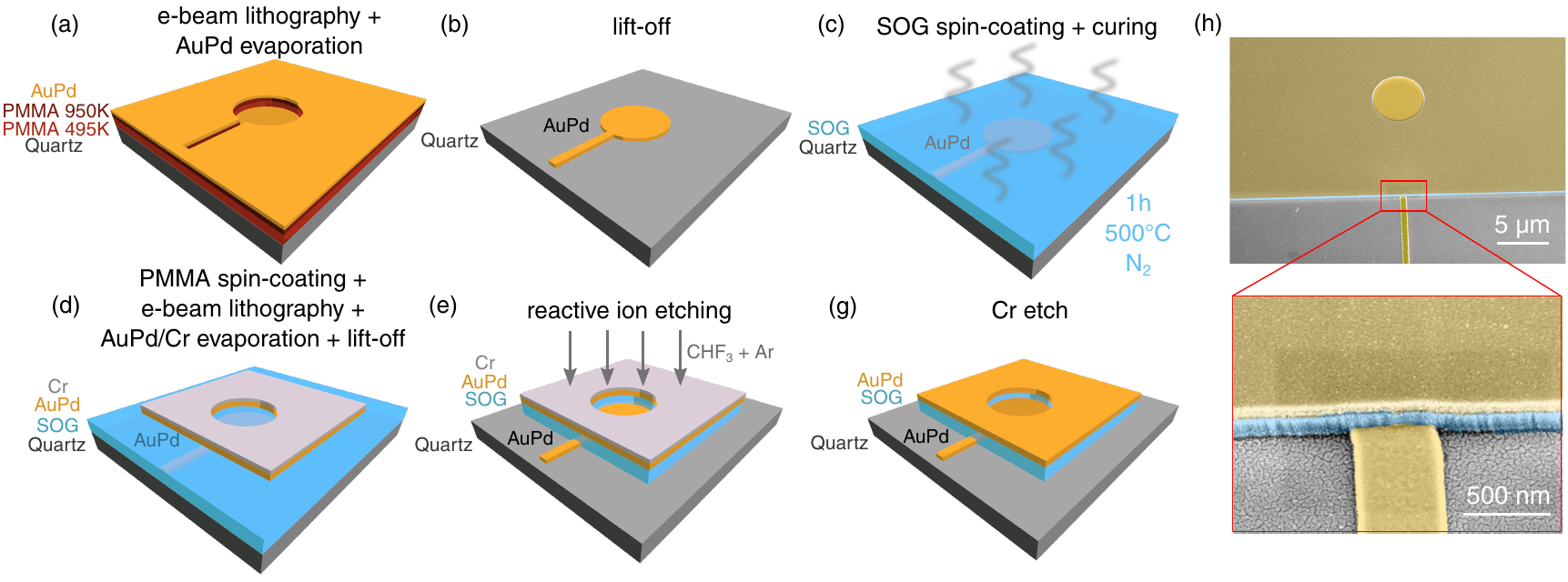}
	\caption{\label{fig:Fig2} (a) - (g) Fabrication steps. (h) A false-colored SEM image of the device showing the top and bottom electrodes (yellow) and the separating SOG layer (blue). A zoomed-in image of the AuPd/SOG/AuPd interface is shown in the bottom panel.}
\end{figure*}
The total capacitance of a single graphene drum and the underlying electrode in typical sample geometries (circular drum 5-10 $\mu$m in diameter, suspended over a 300 nm-deep cavity), ranges from 0.5 to 2 femtoFarads. A displacement of such a drum of 1 nm would result in a capacitance change of only 2-6 attoFarads. Fabricating readout circuitry sensitive enough to detect such changes is faced with a few challenges. (i)~Very shallow gaps are needed in order to maximize the capacitance of the device; (ii)~Parasitic capacitances between the readout electrodes need to be as small as possible to improve the signal to noise ratio; (iii)~The surface should be flat and adhesive to facilitate the transfer of graphene; (iv)~Additionally, to keep the pressure in the reference cavity constant, the cavity needs to be hermetically sealed by the graphene membrane. To address these challenges, we develop a device with electrical readout fabricated on top of a quartz substrate, which substantially reduces the parasitic capacitance of the electrical circuitry. To demonstrate the sensing concept, we transfer a few-layer graphene flake on top of the device and we use external gas pressure load to deflect the drum, reading out the corresponding change in the capacitance.

A 3D schematic of the proposed device is shown in Fig.~\ref{fig:Fig1}(a). The capacitor consists of a circular electrode on the bottom and a suspended graphene drum on top, forming a sealed cavity. The bottom electrode runs underneath a dielectric layer of spin-on-glass (SOG), which separates it from the top metal electrode. The graphene drum is mechanically supported by the top electrode, which also serves as an electrical contact to the graphene. Figure~\ref{fig:Fig1}(b) shows the sensing principle: when the pressure inside the cavity ($P_\mathrm{in}$) is equal to the outside pressure ($P_\mathrm{out}$), the capacitance of the device is given by the parallel plate capacitor formed by the graphene and the bottom electrode: $C_0$. When the outside pressure is higher than the pressure inside the cavity, this results in a positive pressure difference across the membrane, causing it to bulge downwards, which manifests itself as an increase of the measured capacitance. Conversely, if the pressure inside the cavity is higher than the outside pressure, the drum bulges upwards, resulting in a decrease of the measured capacitance.

Fabrication requires two e-beam lithography steps for the bottom and top electrodes. Both lithographic steps use two layers of PMMA resist (A6 495K [300 nm] and A3 950K [100 nm]) in order to create sloped resist walls which facilitate the lift-off. To minimize charging effects during the e-beam patterning, a 10 nm layer of Au is sputtered on top of the resist prior to the e-beam exposure. The Au layer is removed before developing the resist using KI/\ce{I2} gold etchant. Figure~\ref{fig:Fig2}(a) shows a sketch of the sample after developing the resist in MIBK:IPA (1:3) and evaporating 5 nm of titanium (5 nm) and 60 nm of gold-palladium (Au$_{0.6}$Pd$_{0.4}$) to form the bottom electrode. The titanium is used as a thin adhesion layer and is not shown in the figure. 

\begin{figure}[hb]
	\includegraphics[width=0.45\textwidth]{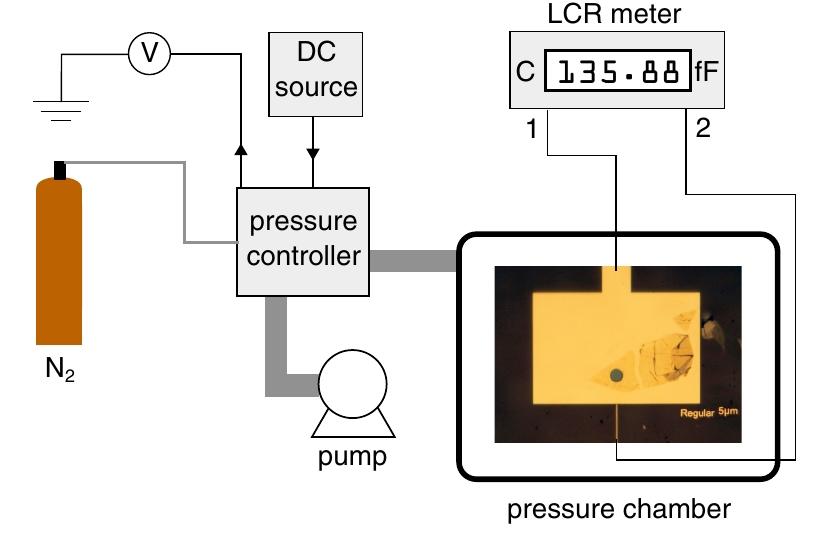}
	\caption{\label{fig:Fig3} Schematic of the measurement setup: the device is mounted in a pressure chamber connected to a pressure controller with a DC voltage control input. The voltage output of the pressure controller is proportional to the actual pressure inside the chamber ($P_\mathrm{out}$). The capacitance of the drum is read out using an LCR meter. The figure shows an optical image of the device.}
\end{figure}

\begin{figure}[h!]
	\includegraphics[width=0.45\textwidth]{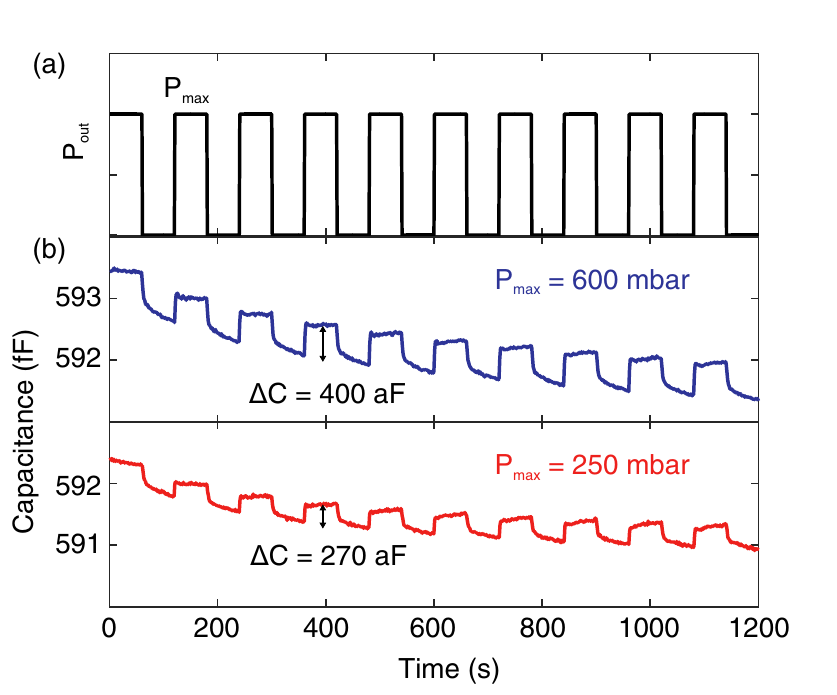}
	\caption{\label{fig:Fig4} (a) Experimental procedure: the pressure is changed from vacuum to $P_\mathrm{max}$ with a period of 120 s. (b) Capacitance of the device as a function of time for two different runs using $P_\mathrm{max} = 600$~mbar (top) and $P_\mathrm{max} = 250$~mbar (bottom). Both panels show the extracted capacitance step height ($\Delta C$) at the moment of changing the pressure.}
\end{figure}
After lift-off (Fig.~\ref{fig:Fig2}(b)), a layer of FOX XR-1451 spin-on-glass (SOG) is spin-coated on the chip. In order to improve the conformity of the SOG layer to the underlying surface, the SOG layer is baked in two stages: 3 minutes at 150 $^{\circ}$C and 3 minutes at 250 $^{\circ}$C. Subsequently, the chip is placed in a \ce{N2} furnace at 500 $^{\circ}$C at 1 atm, which cures the SOG, making it mechanically harder and also improving its surface smoothness and step coverage (Fig.~\ref{fig:Fig2}(c)). The baking and curing processes are essential for obtaining a flat and smooth surface, which is important, as it largely influences the roughness of the electrode evaporated on top of it. Smooth surfaces enhance adhesion and thereby facilitate the transfer of graphene. The current process flow results in a cavity depth of 110 nm. The top electrodes are fabricated on top of the SOG layer, following the same steps of Fig.~\ref{fig:Fig2}(a-b), with a different combination of metals: Ti/$\mathrm{Au_{0.6}Pd_{0.4}}$/Cr (5 nm/90 nm/30). This is shown in Fig.~\ref{fig:Fig2}(d). The top layer of chromium is used as a hard mask for the following etching step, to avoid contamination of the underlying AuPd. \\
\indent Fig.~\ref{fig:Fig2}(e) shows the formation of the cavities by using reactive ion etching (RIE) of the SOG everywhere around the top electrodes. This is done at 7 $\mu$bar in \ce{CHF3}:\ce{Ar} (50:2.5 sccm). The remaining Cr is then etched away using Cr etchant, which results in the final device (Fig.~\ref{fig:Fig2} (g)). The cavity depth can be easily tuned by changing the thickness of the top layer of AuPd. In Fig.~\ref{fig:Fig2}(h) we show a false-colored SEM image of the device after the removal of the Cr. The bottom panel shows a zoom-in of the interface between the two electrodes (yellow) and the SOG layer in between (blue). After the device has been fabricated, graphene flakes are transferred on top of the cavities using a dry transfer technique. The resulting graphene drums are 5~$\mu$m in diameter.\\

The measurement setup is shown in Fig.~\ref{fig:Fig3}. The device is mounted in a vacuum chamber connected to a membrane pump and a pressure controller. The pressure controller is connected to a \ce{N2} gas bottle (purity 99.999~\%) and the pressure of the gas inside the chamber can be controlled linearly by using a 0-10 V input voltage. The pressure controller has a voltage output, which enables a direct readout of the pressure inside the chamber. In this configuration, the pressure can be regulated between 1-1000~mbar (0-10~Volts on the input) with a resolution of $\approx$~0.5~mbar. The capacitance of the graphene drum is measured using an LCR meter in a two-port configuration. All capacitance measurements are performed at a frequency of 1 MHz with a voltage amplitude of $V_\mathrm{p} = 100~\mathrm{mV}$. The integration time for the capacitance readout is 1500 ms. The inset of Fig.~\ref{fig:Fig3} shows an optical image of the measured device: a 6 nm - thick graphene drum.

\begin{figure}[h]
	\includegraphics[width=0.45\textwidth]{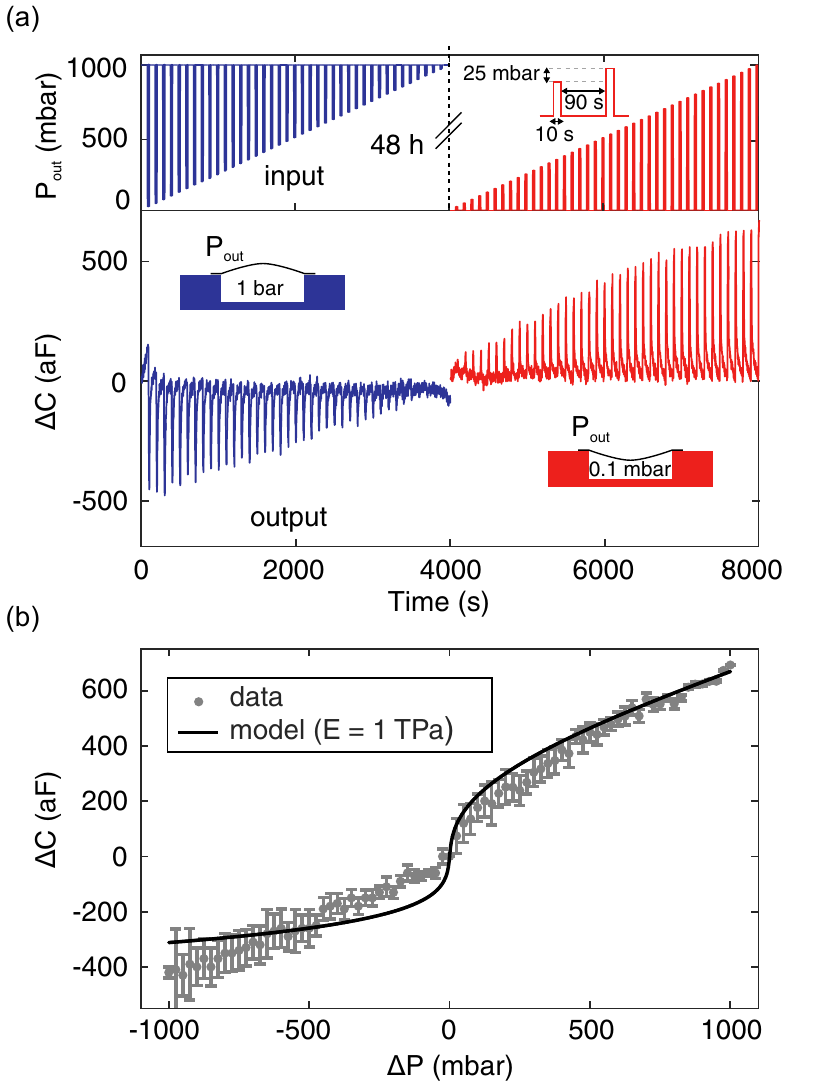}
	\caption{\label{fig:Fig5} (a) Capacitance change as a function of time (top). Bottom: Starting from 1 bar, pressure is changed in a stepwise fashion with increasing steps of 25 mbar (10 \% pulse duration with a 100 s period). The chamber is then pumped down to 1 mbar and a similar procedure is repeated with the chamber being pumped to vacuum after each step. (b) The extracted values for $\Delta C$ as a function of the applied pressure difference across the membrane (gray dots). A theoretical fit to the data using the device dimensions, with pre-tension $n_0 = $ 0.1 N/m and Young's modulus E = 1 TPa.}
\end{figure}


The measurement scheme is sketched in Fig.~\ref{fig:Fig4}(a). Although graphene hermetically seals off the cavity, slow gas permeation usually takes place through its edges or through the underlying oxide~\cite{bunch08}. We make use of this observation and keep the sample in vacuum for 48 h prior to each measurement to ensure that the gas from the cavity is completely evacuated ($P_\mathrm{in}\approx 0$). Then a square wave is applied to the control input of the pressure controller, such that the pressure in the chamber ($P_\mathrm{out}$) is changed in a step-like fashion. Fig. ~\ref{fig:Fig4}(b) shows the measured capacitance of the device as a function of time for two values of the pressure step height: $P_\mathrm{max}~=~$~600~mbar (blue) and 250 mbar (red). Both graphs show that the capacitance rises when the pressure inside the chamber $P_\mathrm{out}$ increases, and jumps back to the initial value upon pumping down. Despite the care taken to eliminate parasitic capacitances, by using a quartz substrate and local gate, the total capacitance of the device is $\approx$ 590 fF, mostly stemming from the parasitic capacitance of the wiring and the on-chip inter-electrode capacitance, since the contribution of the graphene drum is calculated to be only 1.58~fF. 

Starting from $P_\mathrm{in} = P_\mathrm{out}$ and assuming an abrupt change in $P_\mathrm{out}$, such that permeation effects can be neglected, the expected capacitance change can be calculated using an implicit relation between the pressure difference across the membrane ($\Delta P$) and the deflection of the membrane's center ($z$)~\cite{bunch08thesis}:
\begin{equation}
\label{deltaP}
\Delta P = \frac{4n_0}{R^2}z + \frac{8Eh}{3R^4(1-\nu)}z^3,
\end{equation}
where $n_0$ is the pre-tension of the membrane, $R$ and $h$ are its radius and thickness respectively and $E$ and $\nu$ are the Young's modulus and the Poisson's ratio of the material. Knowing $z$ and the spherical deformation shape of the membrane ($U(r) = z(1-\frac{r^2}{R^2})$), the capacitance can be calculated using the parallel plate approximation as:
\begin{equation}
\label{deltaP}
C = 2\pi\varepsilon_0\int_{0}^{R}\frac{r}{g_0 - U(r)}dr,
\end{equation}
where $\varepsilon_0$ is the vacuum permittivity and $g_0$ is the gap size. Using $n_0$ = 0.1 N/m and $E$ = 1 TPa, the value of the extracted capacitance steps matches well with the numbers expected from theory.
On top of the measured signal, we also measure a slow drift of the capacitance over time (see Fig.~\ref{fig:Fig4}). The cause of the drift is not well understood and it might be due to a combination of slow gas leakage and condensation of humidity on the electrodes ~\cite{ford48}.

Using pressure pulses of increasing height we can trace out a dependence of the capacitance change on the pressure difference across the membrane. To do so, we employ a measurement protocol sketched in the inset of Fig.~\ref{fig:Fig5}(a). The sample is kept at vacuum and short pressure steps (10~s) are applied to the sample chamber, followed by 90 s of pumping, to ensure that the cavity underneath the graphene is pumped down to vacuum before applying the next pressure step. This way, it can be safely assumed that the height of the pressure step corresponds to the actual pressure difference felt by the graphene membrane. The opposite applies for the left side of the graph (blue curve): the sample is kept at 1 bar and pressure steps of the opposite sign are applied, followed by 90~s of ambient pressure. The measured capacitance change $\Delta C$ is plotted in the bottom panel of Fig.~\ref{fig:Fig5}(a). The aforementioned drift was subtracted for this dataset after fitting it with a polynomial (see Supporting Information Section I).

The capacitance change is recorded as the height of the step in the measured capacitance immediately after applying the pressure pulse. Doing this for the entire span of $\Delta P$ (from -1 to +1 bar) we get a $\Delta C$ vs. $\Delta P$ curve, plotted in Fig.~\ref{fig:Fig5}(b). The error bars at each point correspond to the RMS noise of the signal in the vicinity of the pressure step, as a measure of the uncertainty of the step determination. The black curve is the modeled response of the system for a 6 nm thick graphene membrane with a Young's modulus of 1 TPa. The model is in a good agreement with the measured response, providing further evidence that the signal is indeed coming from the displacement of the membrane. Thanks to the relatively low parasitic capacitances, despite the present drift, capacitance changes down to 50 aF could be distinguished.

The resolution of the measurement setup is limited by the resolution of the LCR meter, which is 10 aF. This corresponds to a pressure resolution of $\approx$ 360 Pa (or 0.36 mbar) for $\Delta P\approx$ 0 and 10.6 kPa (or 106 mbar) for $\Delta P$ = 1 bar. For potential application of such device as a pressure sensor, it is interesting to look at the sensitivity of the device around $\Delta P$ = 0. By design, the sensitivity of the presented device peaks at around 0.1 aF/Pa at 0 mbar pressure difference (see Supporting Information Section II). The root-mean-square (RMS) noise of the measurement setup is 25 aF/$\sqrt{\mathrm{Hz}}$. However, due to the drift present in the measurements, the minimal step height that could be resolved was 50 aF. The relative error of the pressure measurement ranges from 0.6 \% (for $\Delta P \approx$ 1 bar) all the way up to 300 \% for $-100\,\mathrm{mbar} < \Delta P < 100\,\mathrm{mbar}$. The accuracy of the sensor can also be influenced by morphological imperfections  of the membrane itself. 

There are multiple ways to increase the sensitivity of the device: decreasing the thickness of the graphene ($h$), decreasing the pre-tension of the membrane ($n_0$), increasing its radius ($R$), or connecting $N$ such devices in parallel. A detailed analysis of the influence of each parameter on the sensitivity of the device are shown in the Supporting Information Section II. According to the calculations, changing the thickness $h$ does not drastically influence the sensitivity. Increasing $R$ or decreasing $n_0$ improve the sensitivity by one or two orders of magnitude. We note that controlling the pre-tension is challenging, since it largely depends on the transfer process and usually results in large spreads~\cite{hone08elastic}. Moreover, making devices with larger radii and low pre-tension would impair the yield of the devices~\cite{cartamil17} and reduce their dynamic range (due to collapse of the membrane at high $\Delta P$). However, increasing the number of drums in parallel $N$ linearly increases the responsivity of the device. With more than 1000 drums in parallel (resulting roughly in a chip size of 100 x 100 $\mu m$) one could push the responsivity to values higher than 100 aF/Pa, resulting in a 0.1 Pa resolution using our current measurement setup.\\
\indent To demonstrate the feasibility of capacitive readout of the graphene sensor with an integrated circuit, we replaced the LCR meter with an Analog Devices (AD7746) capacitance-to-digital converter chip (with dimensions 5x5 mm$^2$) which we interfaced through using the built-in I2C protocol library of an Arduino. We show an example of such measurement in the Supporting Information Section III. Even though the signal-to-noise of this measurement is worse than the one using the LCR meter, it still serves as a proof-of-principle that on-chip detection of small capacitance changes can be realized using commercial electronic devices.\\
\indent The drift in the measurement together with the poor hermeticity of the membrane hamper the long-term stability of the device. For its commercial application as a pressure sensor, the hermeticity of the device needs to be improved (e.g. by properly sealing the membrane edges) and the cross-sensitivity to the environment (humidity/gas composition) needs to be investigated more thoroughly.\\
\indent In conclusion, we demonstrate on-chip capacitive readout of a single suspended graphene drum. To obtain the responsivity required for sensing such small capacitance changes, the entire fabrication is done on an insulating quartz substrate, minimizing the parasitic capacitance of the readout electrodes. We use uniform pressure load to statically deform the membrane, which results in a capacitance change of the device. Using this method, we are able to read out capacitance changes down to 50 aF and detect pressure steps down to 25 mbar. The height of the steps is consistent with predictions from the theoretical model. We also traced out a force-deflection curve by pulsing the pressure in the chamber with pulses of increasing height. The measured $\Delta C$ vs. $\Delta P$ curve matched well with theory, based on a graphene membrane with a Young's modulus of 1 TPa. We also measured a temporal drift in the capacitance, possibly originating from residual humidity in the chamber. This work is aimed at probing the limit of static capacitive detection of graphene nanodrums. We optimized the device to enable detection of very small capacitance changes of down to 50 aF. By combining this device design with an on-chip capacitance-to-digital converter we show a proof-of-concept demonstration of the feasibility of integrating suspended 2D membranes into next-generation pressure sensors.

\section*{Acknowledgements}

This work was supported by the Netherlands Organisation for Scientific Research (NWO/OCW), as part of the Frontiers of Nanoscience (NanoFront) program and the European Union Seventh Framework Programme under grant agreement $\mathrm{n{^\circ}~604391}$ Graphene Flagship.\\

\pagebreak
\onecolumngrid
\setcounter{figure}{0}
\renewcommand{\figurename}{FIG. S\!\!}

\section*{Supporting Information}
\subsection*{I. Background drift subtraction}
\begin{figure}[H]
	\begin{center}
	\includegraphics[width=0.45\textwidth]{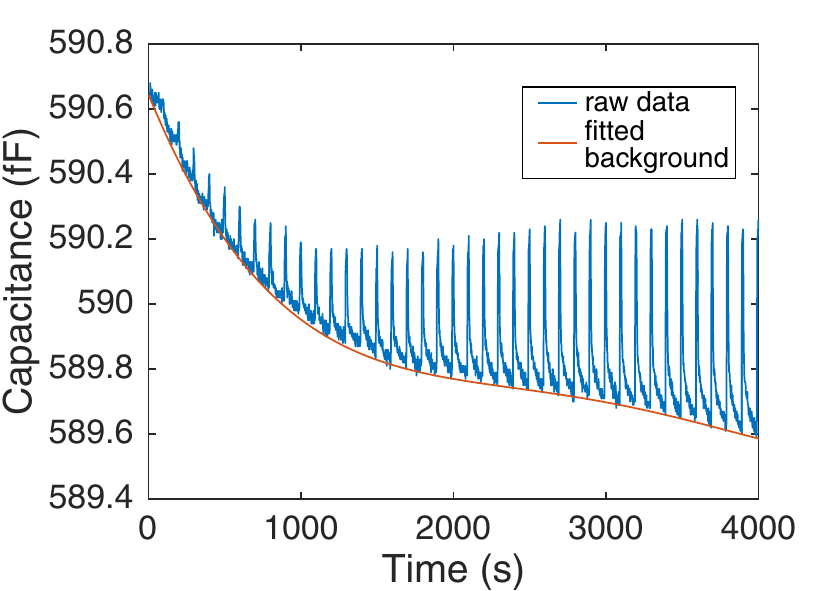}
	\caption{\label{fig:FigS1} Drift subtraction: the blue curve represents the raw capacitance data, while the red curve is a polynomial fit of the local minima of the capacitance after each pumping step.}		\end{center}
\end{figure}

In this section we present an example of drift subtraction using the raw data of the $\Delta C$ vs. $\Delta P$ measurement presented in Fig. 5 (a) of the main text (for $P_\mathrm{in} \approx 0$ mbar). In Fig.~S\ref{fig:FigS1} the measured capacitance signal is plotted against time. As described in the main text, the pressure is increased in 25 mbar steps (with 10 s duration) followed by 90 s of pumping the chamber to vacuum. We can distinguish a background signal (defined by the minima of the measured capacitance after each pumping step) on top of which we measure sharp steps in the capacitance at the moment of introducing gas inside the chamber. This drift can be subtracted from the measurement by fitting a (4$^\mathrm{th}$ order) polynomial through these minima (red line) and subtracting it from the data (blue line). The resulting signal represents the $\Delta C$ vs. $\Delta P$ curve plotted on the right-hand side of Fig. 5 (a) of the main text.\\
\newpage

\subsection*{II. Pressure sensitivity}
In this section we lay out the theoretical predictions of the expected capacitance change as a function of the pressure differences across the membrane, together with the expected responsivity of the device.

Using eqs. (1) and (2) from the main text we model the predicted capacitance response as a function of the pressure difference across the membrane. The pressure difference $\Delta P$ is defined as $\Delta P = P_\mathrm{out} - P_\mathrm{in}$, where $P_\mathrm{out}$ is the pressure in the chamber and $P_\mathrm{in}$ is the pressure inside the cavity.

We start with the dimensions of the device described in the main text, namely, a single graphene membrane with a thickness of $h = 6$ nm, pre-tension $n_0=0.1 N/m$ and a radius $R = 2.5~\mu$m. For each of the panels (a-d) we keep all parameters fixed, while varying only (a) the thickness, (b) the pre-tension, (c) the radius and (d) the number of identical drums connected in parallel. The top panels represent the absolute value of the expected capacitance change ($|\Delta C|$) for a given pressure difference ($\Delta P$). The bottom panels represent the calculated responsivity ($\frac{\partial C}{\partial \Delta P}$) of the device, expressed in aF/Pa.

\begin{figure}[H]
	\begin{center}
	\includegraphics[width=1\textwidth]{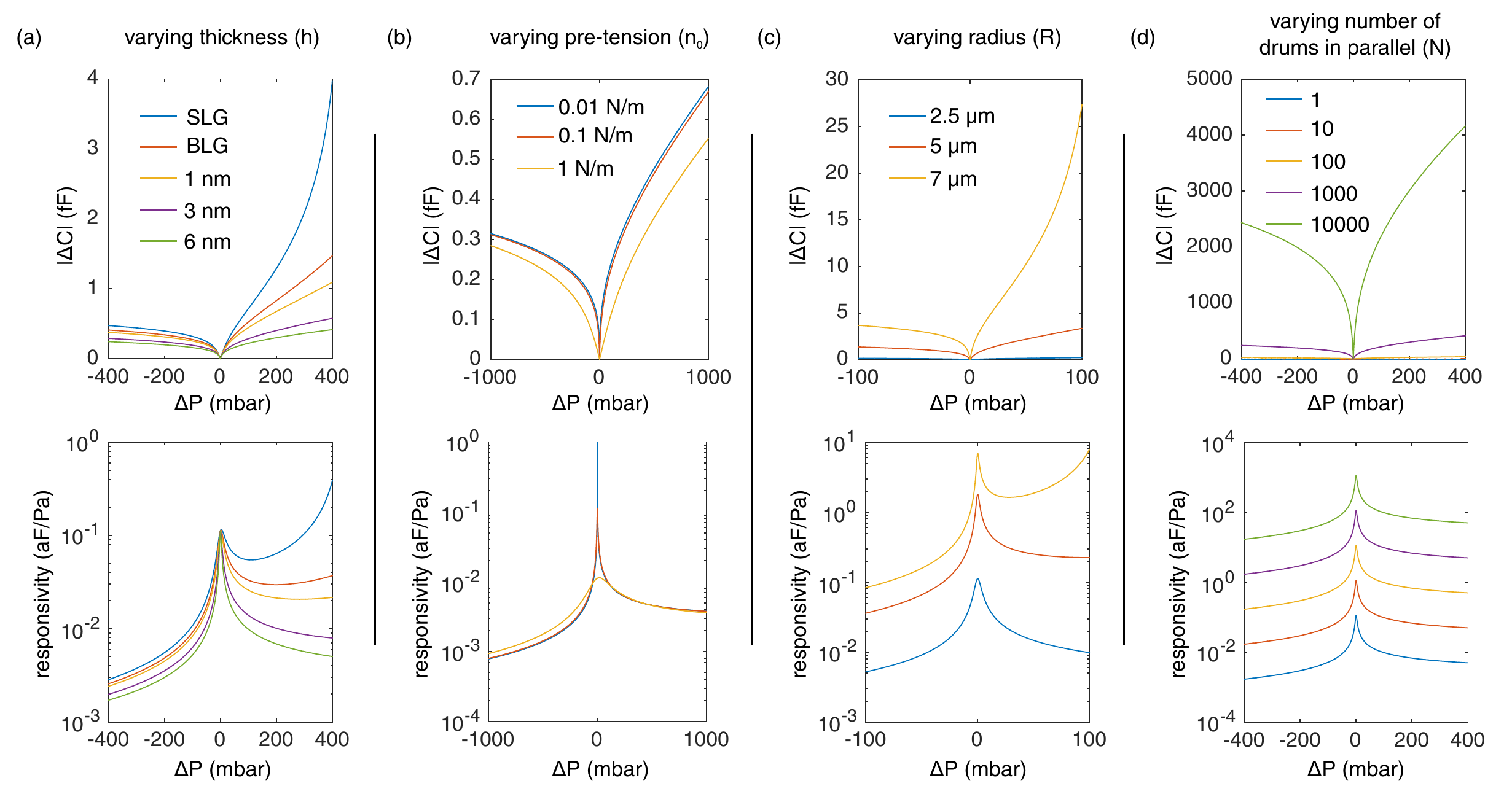}
	\caption{\label{fig:FigS2} Calculated capacitance change (top panels) and responsivity (bottom panels) as a function of pressure difference for varying (a) thickness, (b) pre-tension, (c) radius of the membrane and (d) number of graphene drums in parallel.}
	\end{center}
\end{figure}
To be able to use such a device as a pressure sensor in mobile devices, considering the day-to-day fluctuations in the atmospheric pressure, one needs to look at the -100 mbar $\le \Delta P\le$ 100 mbar region of the responsivity plots, where the responsivity is usually the highest.

It can be seen from the plots that by varying the membrane's thickness or pre-tension, one does not benefit a lot when it comes to the responsivity of the device. A larger radius, which results in a higher initial capacitance ($C_0$), naturally, increases the overall responsivity of the device. It has to be noted, however, that making the membranes larger, thinner or with a lower pre-tension also significantly reduces the dynamic range of the device, because it makes it easier for the membrane to collapse at a high pressure difference $\Delta P$. Mathematically this follows directly from eq. 1 of the main text. 

One solution for increasing the responsivity of the device is increasing the number of drums connected in parallel (Fig.~S\ref{fig:FigS2} (d)). This way, the dynamic range of the sensor is unchanged, whereas the overall responsivity of the system scales proportionally to the number of drums $N$. Using 1000 drums in parallel (resulting roughly in a 100 x 100 $\mu m^2$ device), the responsivity peaks at 100 aF/Pa, which would, with a capacitance resolution of 10 aF, enable a pressure resolution of 0.1 Pa. Such a device could be used in next-generation pressure sensors and could have a comparable or better performance than the current state-of-the-art pressure sensors.

\newpage
\subsection*{III. Readout using an AD7746 chip interfaced with an Arduino}
To demonstrate the integrability of the proposed device, we replaced the LCR meter with a compact, low-cost
Analog Devices (AD7746) 24-bit capacitance-to-digital converter chip. According to the specifications, this chip is able to handle up to 4 pF parasitic capacitance and has a resolution of down to 4 aF. The dimensions of the readout chip are $\approx$ 5 x 5 mm$^2$ and it has a built-in I$^2$C interface. 
\begin{figure}[H]
	\begin{center}
	\includegraphics[width=0.8\textwidth]{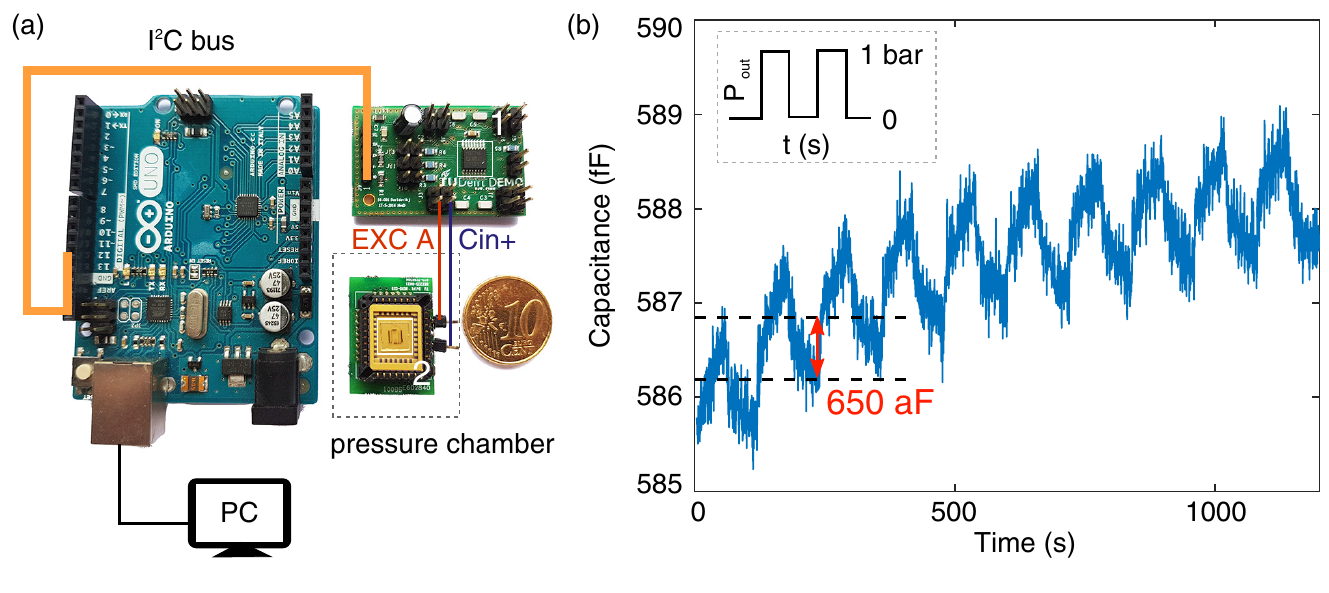}
	\caption{\label{fig:FigS3} (a) Alternative measurement setup using an AD7746 capacitance-to-digital converter chip (1). The AD7746 is connected to the sample (2) using two wires (EXC A and Cin+). The sample together with the chip carrier is mounted inside the pressure chamber. The AD chip is interfaced with I$^2$C communication using an Arduino Uno, which is connected directly to the measurement computer to read out the capacitance of the device. The 10 cent euro coin is shown for scale (20 mm in diameter). (b) Capacitance signal measured using the AD7746 chip following the same measurement scheme from Fig. 4 in the main text. The pressure step is 1 bar and the corresponding capacitance step is 650 aF, consistent with theory.}
	\end{center}
\end{figure}

The entire measurement setup (without the pressure chamber and the pressure controller) is shown in Fig.~S\ref{fig:FigS3} (a). The AD chip is mounted on a printed circuit board (PCB) together with a few resistors and capacitors, necessary for its basic operation. Two wires from the chip (EXC A (red) and Cin+ (blue)) are connected directly to the top and bottom electrode of our device respectively. Our entire chip (marked with (2) in the image) is mounted on a chip carrier inside the pressure chamber. The AD chip is interfaced using an Arduino Uno, which has a built-in I$^2$C protocol library. The Arduino is connected to a measurement computer, which records the capacitance value measured by the AD7746 chip.\\
To test the setup, we employ a measurement scheme similar to the one shown in Fig. 4 of the main text. After keeping the device for 48 hours in vacuum, we apply 1 bar pressure steps with a duration of 60 s, followed by 60 s of pumping. The measured capacitance signal is shown in Fig.~S\ref{fig:FigS3} (b). We observe a capacitance change of $\Delta C\approx 650$ aF, which is consistent with the measurement using the LCR meter for $\Delta P = 1$ bar (see Fig. 5 of the main text). The noise level of these measurements is much higher (79 aF/$\sqrt{\mathrm{Hz}}$ RMS), as is the observed drift of the capacitance signal.\\
Part of the reason for the increased noise is the faster sampling time (109 ms), but also the fact that the PCB with the AD7746 chip was kept outside of the vacuum chamber (due to pressure sensitive capacitors on the PCB) and long unshielded wires were used to connect to the sample. This could be improved by redesigning the PCB and wire-bonding the graphene device directly to the PCB, in the proximity of the AD readout chip. Nevertheless, this is a proof-of-concept measurement, showing that our graphene device can be used as a pressure sensor with an all-electrical on-chip readout solution.

\end{document}